 \documentclass[doublecol]{epl2} 

 \usepackage{amsmath}
\usepackage{color}
\usepackage{latexsym}
\usepackage{amssymb}

\title{Wetting of crossed fibers: multiple steady states and symmetry breaking}

\author{Alban Sauret\inst{1} \and Alison D. Bick\inst{1} \and Camille Duprat\inst{2} \and Howard A. Stone\inst{1}}
\shortauthor{A. Sauret \etal}

\institute{                    
  \inst{1} Department of Mechanical and Aerospace Engineering, Princeton University, Princeton, New Jersey 08544, United States\\
  \inst{2} Laboratoire d'Hydrodynamique (LadHyX) and Department of Mechanics, Ecole Polytechnique, CNRS, 91128 Palaiseau Cedex, France
}

\abstract{We investigate the wetting properties of the simplest element of an array of random fibers: two rigid fibers crossing with an inclination angle and in contact with a droplet of a perfectly wetting liquid. We show experimentally that the liquid adopts different morphologies when the inclination angle is increased: a column shape, a mixed morphology state where a drop lies at the end of a column, or a drop centered at the node. An analytical model is provided that predicts the wetting length as well as the presence of a non-symmetric state in the mixed morphology regime. The model also highlights a symmetry breaking at the transition between the column state and the mixed morphology. The possibility to tune the morphology of the liquid could have important implications for drying processes.}

\begin{document}

\maketitle

\section{Introduction}

Fibrous media are common in various engineered systems such as filters, paper or the textile industry. Many of these materials can be described as networks of disordered fibers. For instance, fibrous filters exhibit many locations where the fibers cross with various angles \cite{last,contal2004,hubbe2008}. The presence of a wetting liquid, due to the ambient humidity or purposefully added, can modify the internal properties of the materials as the liquid typically spreads around the points where the fibers cross. 

The effects of a wetting liquid have been studied in detail for wet granular material, such as sand, where the moisture has been shown to modify the rheological properties of the bulk material \cite{stone2001,tregzes2002,nowak2005,herminghaus2005,kudrolli2008,chopin2011,bonn2012,strauch2012}. In this case, the interaction between two spherical beads is modeled by a liquid bridge \cite{willett2000}. However, the situation is different with anisotropic materials where the morphology of the liquid is more complex because of the geometrical constraints. This situation is more complicated to describe, as the influence of humidity on such media needs a model different from spherical beads, but is nevertheless significant since recent studies characterize flows of dry anisotropic media \cite{rod1,rod2,rod3,rod4}.

Therefore, it is important to understand the morphology of liquid on a pair of rigid fibers. To date, most wetting studies have considered a single fiber \cite{caroll1976,caroll1986,chou2011} or parallel fibers, both rigid \cite{princen1970,bedarkar2010,protiere2013} and flexible \cite{bico2004,duprat2012nature,duprat2013}, where two distinct morphologies, e.g. an extended column state or a drop shape, are possible. The results of such models can not be applied directly to common cases such as an array of randomly oriented fibers where different morphologies have been observed \cite{mullins,claussen}. Thus, there is a need to model and understand the morphology of the wetting liquid in crossed fiber systems, which is important to characterize a large array of fibers and make possible a model of wet anisotropic granular media. Moreover, the ideas are applicable to digital microfluidics on fibers \cite{gilet2009}, which needs an understanding of the wetting at the nodes of two crossed fibers. For instance, tuning the angle between the fibers could lead to variation in the volume collected when a droplet slides along a fiber. 

In this Letter, we consider the simplest element of a random array of fibers: two crossed rigid fibers clamped at both ends. In addition, we also model a fiber as a straight cylinder of constant cross-section. Through a systematic experimental study, we characterize the different possible morphologies and we identify a new composite drop and column state that we denote a mixed morphology state. We also provide an analytical model that allows us to quantify the length of the column shape as well as the transition between mixed and column morphologies including the symmetry breaking transition we have documented.


\section{Experimental setup and observations}

\begin{figure}
\begin{center}
\includegraphics[width=8.8cm]{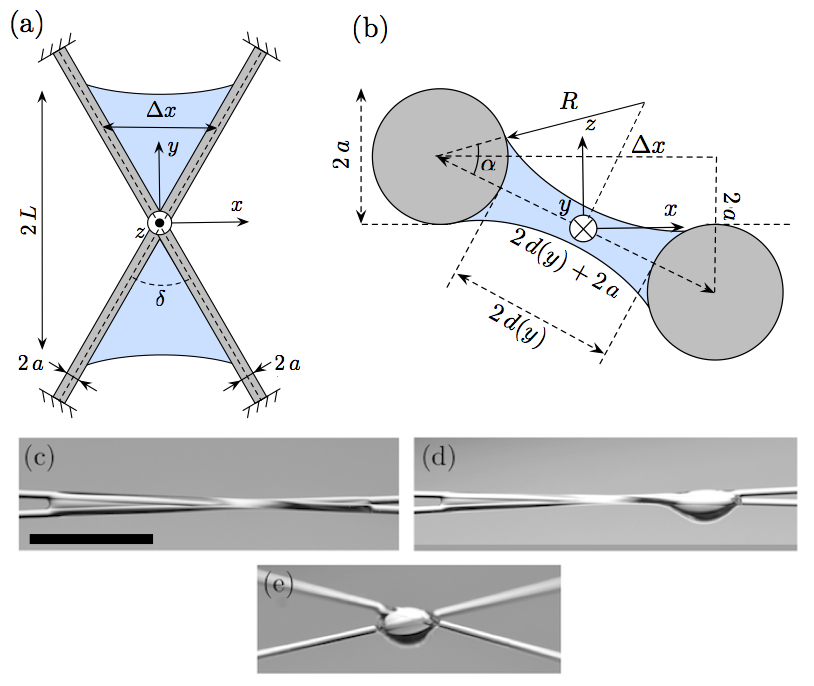}
\caption{Schematic of the wetting of the crossed fibers: (a) top and (b) front views. A liquid can adopt (c) a column shape, (d) a {mixed morphology state} or (e) a drop shape centered at the fiber crossing. Scale bar is $5$ mm.}\label{figure1}
\end{center}
\end{figure}

We deposit a small amount of perfectly wetting liquid (5 cSt silicone oil, contact angle $\theta_e=0^{\circ}$) between two fibers crossed at an angle $\delta$ (Fig. \ref{figure1}). Typically, $\delta$ lies in the range $[0.5^{\circ},30^{\circ}]$, whereas the volume of liquid is $V =1-8 \,\mu$L. We used nylon fibers of various diameters {$2\,a=0.20-0.35$ mm}. In the present study, gravitational effects are neglected since the capillary length $\ell_c=\sqrt{\gamma/(\rho\,g)}$ is about 1.5 mm, which is larger than the typical height $H$ of the liquid in our experiments. A schematic of the fibers and the wetting morphologies are shown in Fig. \ref{figure1} together with the notations we use. Provided that the fibers are not parallel, i.e. they have a small inclination angle $\delta$ between them, a drop of liquid bridging two fibers will travel towards the point where the fibers are the closest, which we refer to as the fiber crossing if the fibers are in contact (see supplementary material SI). This movement is a result of a pressure difference between the two sides of the drop or the column, which is similar to the behavior of a drop on a conical fiber \cite{quere_conical}. The drop then reaches a stationary state; here we only consider this static equilibrium configuration. Depending on the volume of liquid and the angle $\delta$ between the fibers, three distinct equilibrium shapes are observed: (i) a long liquid column \cite{protiere2013} (Fig. \ref{figure1}(c)), (ii) a {mixed morphology state} that consists of a drop on one side together with a small amount of liquid on the other side (Fig. \ref{figure1}(d)), or (iii) a drop or a compact hemispherical drop (Fig. \ref{figure1}(e)). In contrast to studies with rigid parallel fibers, the cross-sectional area of the liquid is not constant along the column as the inter-fiber distance $d(y)$ is varying.

\begin{figure}
\begin{center}
\includegraphics[width=8.7cm]{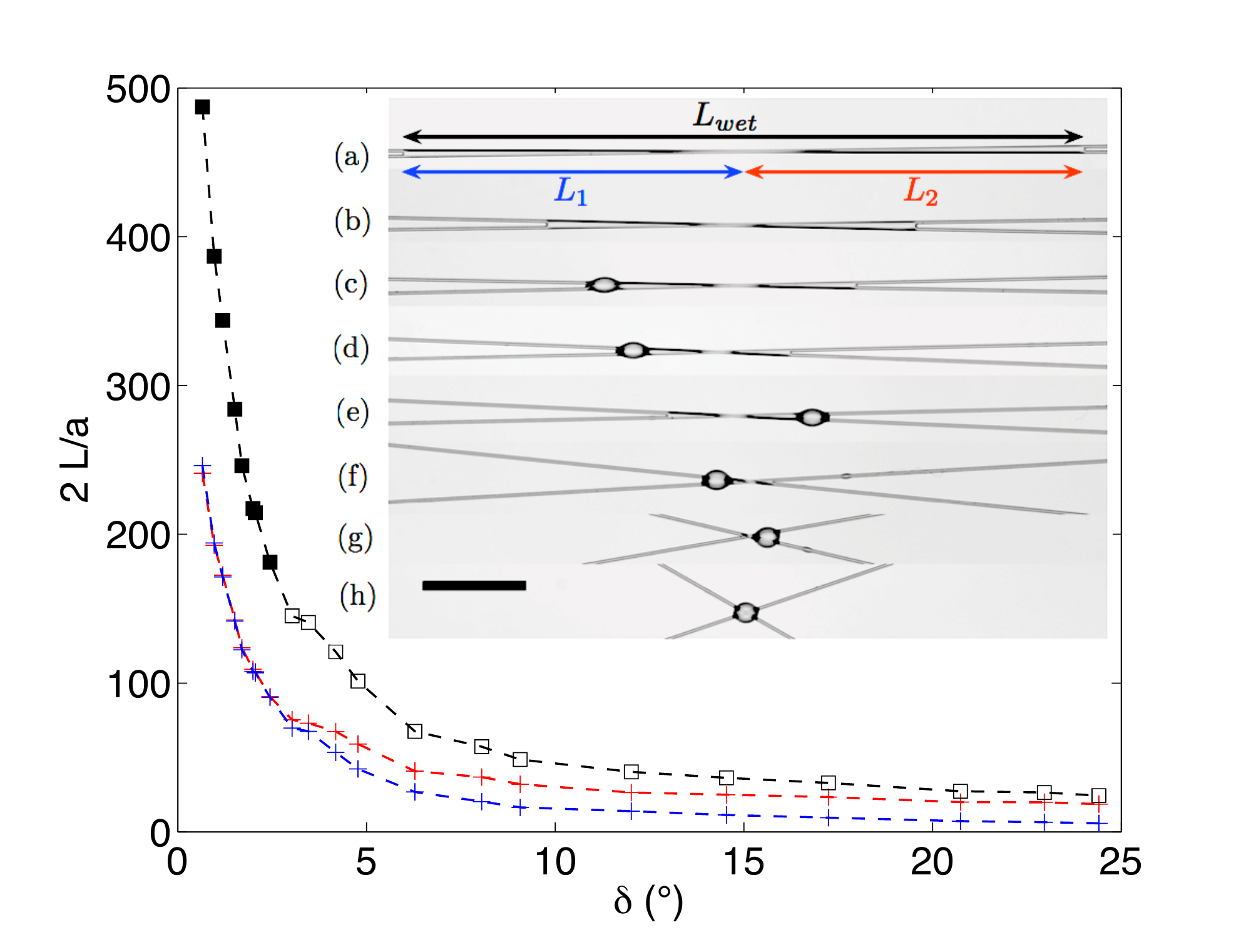}
\caption{Evolution of the length of the liquid when the tilting angle between the two fibers $\delta$ is varied. Nylon fibers of diameter $2\,a=0.30$ mm and liquid volume $V/a^3=1000$. The black symbols indicate the total wetting length $2\,L/a$ (filled: column shape, empty: {mixed morphology}), the blue and red crosses denote the length on both sides of the nodes $L_1/a$ and $L_2/a$. Inset: snapshots of the morphology for increasing angles (a) $\delta=0.9^{\circ}$, (b) $\delta=2.0^{\circ}$, (c) $\delta=2.3^{\circ}$, (d) $\delta=3.6^{\circ}$, (e) $\delta=4.0^{\circ}$, (f) $\delta=8.8^{\circ}$, (g) $\delta=23.1^{\circ}$ and (h) $\delta=49.3^{\circ}$. Scale bar is $10$ mm.}\label{figure2}
\end{center}  
\end{figure}

Typical results are shown in Fig. \ref{figure2} where we measured the wetting length as a function of the tilt angle $\delta$. The wetting length decreases while increasing $\delta$. Above a critical angle, we observe a transition to a {mixed morphology} state, which is characterized by a bifurcation from a symmetric state where both sides have the same liquid length to a non-symmetric state with one drop on one end of the column for any initial condition. A symmetric {mixed morphology} state with a drop on both sides is only a transient state and is not an equilibrium state (see supplementary material SII).

\begin{figure}
\begin{center}
\includegraphics[width=8.5cm]{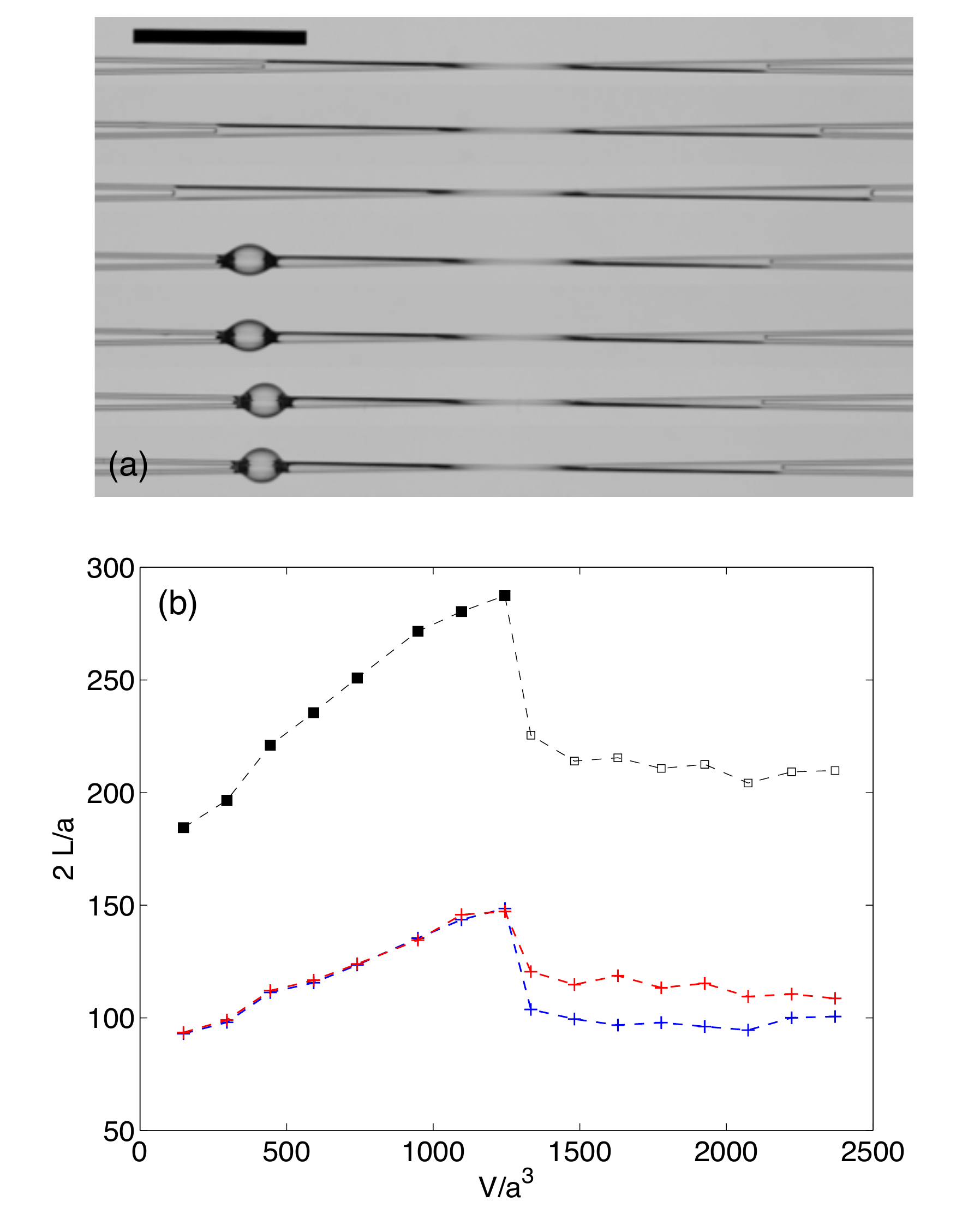}
\caption{{(a) Snapshots of the morphology observed for increasing volumes and for nylon fibers of diameter $2\,a = 0.30$ mm and a constant tilting angle $\delta=1.7^{\circ}$. From top to bottom $V/a^3=296$, $593$, $948$, $1481$, $1778$, $2074$ and $2370$. Scale bar is $10$ mm. (b) Corresponding evolution of the wetting length $2\,L/a$ as the volume of fluid $V/a^3$ is increased. The black symbols indicate the total wetting length $2\,L/a$ (filled: column shape, empty: mixed morphology), while the blue and red crosses denote the length on both sides of the nodes $L_1/a$ and $L_2/a$.}}\label{volume}
\end{center}
\end{figure}

{The influence of the volume on the observed morphology is illustrated in Fig. 3. We measured the wetting length for increasing volume at a fixed angle $\delta$ and fiber radius $a$. We observe that the length of the column initially increases with the volume. Above a critical threshold the column retracts, which leads to a mixed morphology as observed in Fig. 2. A further increase in volume does not necessarily lead to a drop state. In particular, our experiments suggest that the drop morphology depends mainly on the angle $\delta$ between the two fibers at sufficiently large volume and is not observed below a critical angle. The characterization of this second transition is more complicated to determine with our experimental setup and will be the subject of future studies.}

\begin{figure}
\begin{center}
\includegraphics[width=8.8cm]{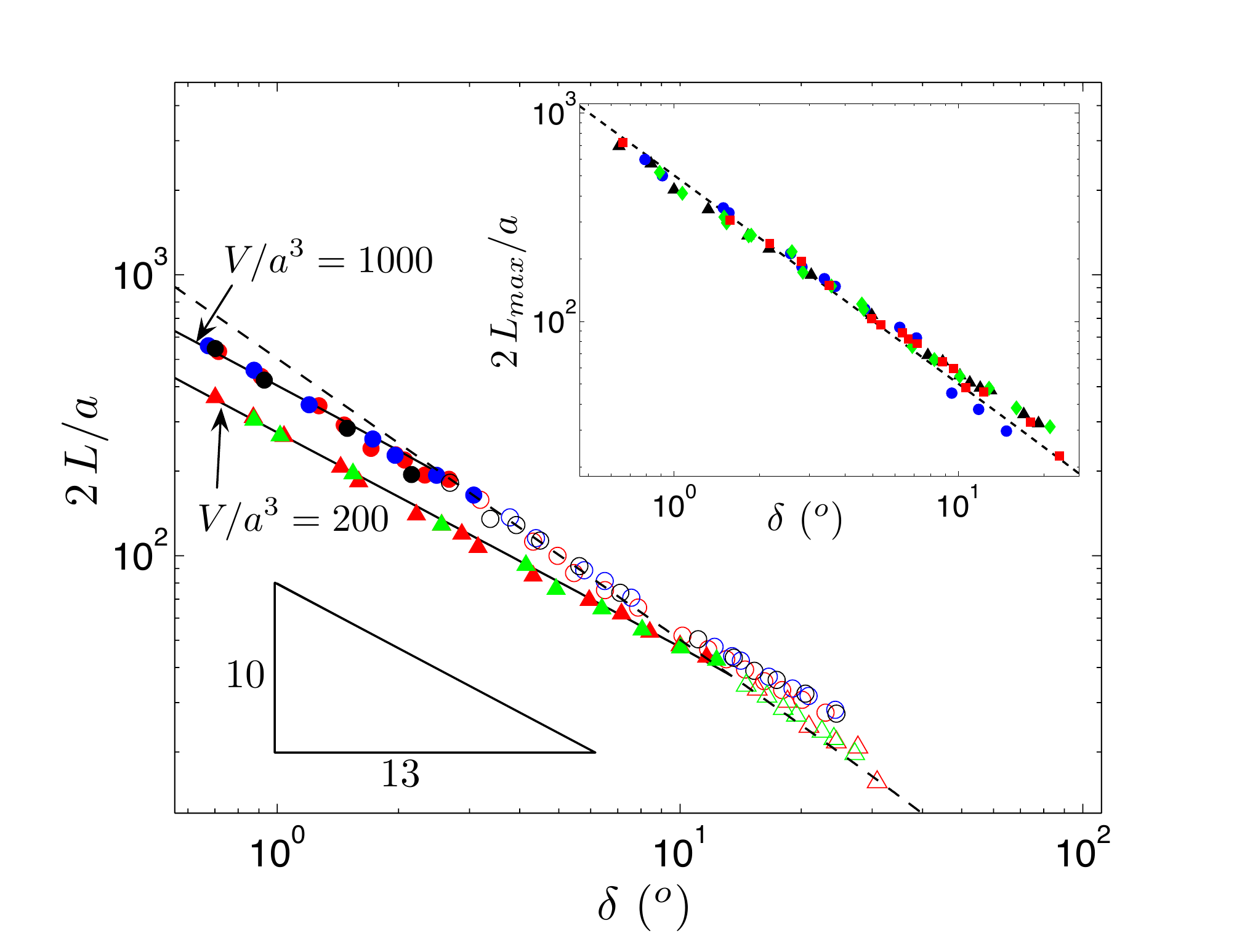}
\caption{Evolution of the wetting length $2\,L/a$ as the tilting angle $\delta$ is varied for nylon fibers of diameter $2\,a = 0.20$ mm (black), $2\,a=0.245$ mm (blue), $2\,a=0.30$ mm (green)  and $2\,a=0.35$ mm (red) and two different volumes: $V/a^3=1000$ (circles) and $V/a^3=250$ (triangles). {The solid symbols are obtained in the column morphology and are therefore well captured by the analytical expression whereas the open symbols are in the mixed morphology and are not predicted by the column state.} The continuous lines are the analytical predictions, the dashed line is the maximum wetting length given by (\ref{numero2}) and the triangle shows the asymptotic scaling law, $L \propto \delta^{-10/13}$. Inset: maximum wetting length $L_{max}(\delta)$ obtained by increasing the volume until the column to mixed morphology transition occurs. {Note that these data represents various volumes of fluid.} }\label{figure3}
\end{center}
\end{figure}


\section{Modelling}

In contrast to studies devoted to parallel fibers \cite{princen1970,protiere2013}, the inter-fiber distance $d(y)$ depends on $y$, the distance to the point where the two fibers cross. From geometric considerations (see Fig. \ref{figure1}), we have
\begin{equation}\label{toto1}
\frac{d(y)}{a}=\sqrt{\frac{y^2}{a^2}\,\tan^2\left(\frac{\delta}{2}\right)+1}-1.
\end{equation}
We also obtain the radius of curvature of the interface ${R(y)}=(a+d(y))/\cos \left[\alpha(y)\right]-a$. We assume that at each location along the column the shape of the liquid cross-section only depends on the rescaled distance between the fibers, and we neglect the detailed shape of the meniscus at the ends as proposed by Princen \cite{princen1970}. We assume a variation of liquid length with the cross-sectional area $A(y)$ of the liquid column such that
\begin{equation}\label{2}
\frac{A}{a^2}=\frac{R^2}{a^2}\left(2\alpha-\pi+\sin(2\alpha)\right)+\frac{2R}{a}\,\sin(2\alpha)-2\alpha+\sin(2\alpha),
\end{equation}
where $\alpha(y)$ is the half-opening angle at the liquid-fiber-air interface (see Fig. \ref{figure1}). An energy balance on a small liquid volume $\text{d}V=A\,\text{d}L$ leads to
\begin{equation}\label{1}
4\,\left[\left(\frac{\pi}{2}-\alpha(y)\right)\,R(y)-\alpha(y)\,r\right]=-\frac{A(y)}{R(y)},
\end{equation}
where the left-hand side is proportional to the change in interfacial energy during the virtual displacement $\text{d}L$ (the first term being proportional to the capillary force integrated at the air-liquid interface, the second term being proportional to the force exerted by the fiber on the drop). The right-hand side is proportional to the work against the Laplace pressure. In this relation, we have assumed that the interface is in local equilibrium at every cross-section and that the interface curvature along $y$ can be neglected with respect to the Laplace curvature of a cross section. Inserting (\ref{2}) into (\ref{1}) leads to a quadratic equation for $R(y)$, whose solution is: 
\begin{equation}\label{4}
\frac{R(y)}{a}=\left[\sqrt{\frac{\pi}{2\,\alpha(y)-\sin[2\,\alpha(y)]}}-1\right]^{-1}.
\end{equation}
Finally,  we obtain
\begin{equation}\label{5} 
\sqrt{\frac{y^2}{a^2}\,\tan^2\left(\frac{\delta}{2}\right)+1}=\left(1+\left(\frac{R(y)}{a}\right)^{-1}\right)\,\cos[\alpha(y)].
\end{equation}

The volume of liquid lying on the fibers can be computed numerically with (\ref{2}) and $\alpha(y)$ obtained from the relation (\ref{5}): $V  =  \int_{-L}^{L}\,A(y)\,\text{d}y$. For a given volume $V$ there is a unique value of the wetting length $L$. The comparison of the experimental results for different fiber radii $a$ and two rescaled volumes $V/a^3=1000$ and $V/a^3=250$ are reported in Fig. \ref{figure3}. We observe a near perfect agreement provided that all of the liquid is in a column state.

In addition, we can calculate the maximum length at which the fluid can spread in a column state, since this corresponds to a half-opening angle $\alpha=\pi$. The maximum inter-fiber distance $L_{max}$, corresponding to the column/drop transition, corresponds to $d=\sqrt{2}$ \cite{princen1970,protiere2013}, which leads to
\begin{equation} \label{numero2}
\frac{L_{max}}{a}=\frac{\sqrt{2+2\,\sqrt{2}}}{\tan\left(\delta/2\right)}.
\end{equation}
If the volume of fluid wetting the fiber is such that all of the fluid can not be contained within the distance $L_{max}$, an accumulation of fluid will appear at one end of the column leading to a {mixed morphology} state. The inset of Fig. \ref{figure3} shows the experimental values of this maximum length compared to the prediction (\ref{numero2}). We added liquid to the column until the {mixed morphology state} was reached, which determines the symmetry-breaking transition between the two states and then $L_{max}$. Again, the model predictions exhibit a very good agreement for various fiber diameters. From these results, the length of the column is first given by $L(\delta)$, then the transition to the {mixed morphology state} is reached when $L=L_{max}$. As $L(\delta)$ depends on the volume of fluid, the transition is reached for lower angles at large volumes.

Also, we observe that for small tilting angles, the wetting length follows a power law. Analytically, in the limit of small $\delta$ and $\alpha \ll 1$ (small volume of liquid), the radius of curvature (\ref{4}) reduces to $R/a \sim 2\,\alpha^{3/2}/\sqrt{3\,\pi}$ and the cross-section area (\ref{2}) reduces to $A/a^2 \sim 8\,\alpha^{5/2} /\sqrt{3\,\pi}$. With the expression (\ref{5}) and the integral for the volume of fluid we obtain the wetting length $L \propto \delta^{-10/13}$, which is indicated in Fig. \ref{figure3} and is in good agreement with the data.

\begin{figure}
\begin{center}
\includegraphics[width=8.8cm]{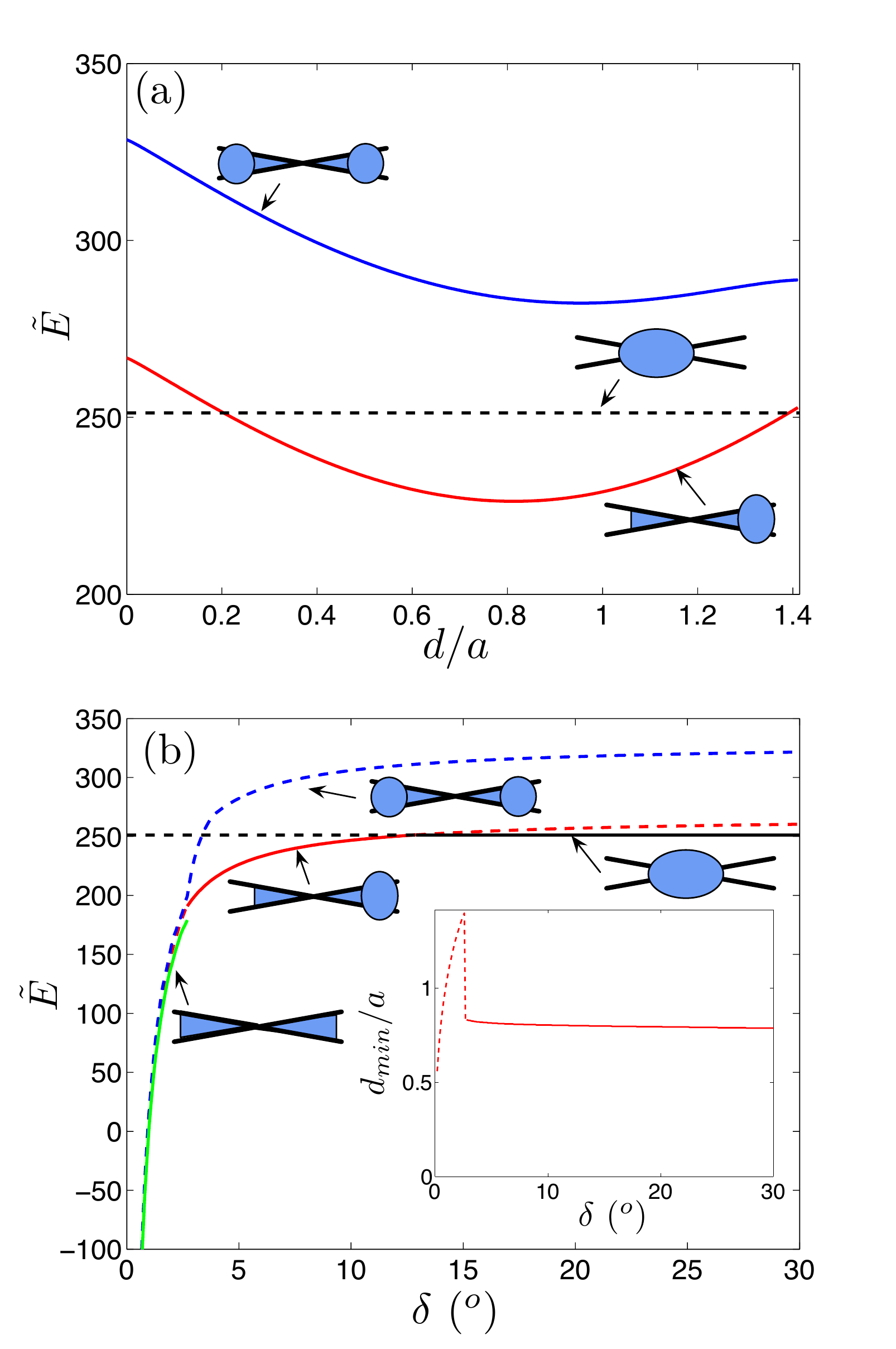}
\caption{(a) Evolution of the surface energy $\tilde{E}=E/(\gamma\,a^2)$ as a function of $d/a$, the inter-fiber distance at the position of the drop, for $V/a^3=1000$ and $\delta=5^{\circ}$ and different morphologies (see schematics). (b) Evolution of the minimum of the surface energy $\tilde{E}$ (with respect to $d/a$) as a function of the tilting angle $\delta$ for $V/a^3=1000$ and different morphologies (see schematics). The continuous line indicates the most stable state at a given $\delta$. Inset: value of $d_{\rm min}/a$ which minimizes the surface energy $\tilde{E}$ for varying $\delta$ {in the mixed morphology}. The dashed line indicates the column {morphology where $d_{\rm min}$ is the inter-fiber distance at the end of the column. The continuous line denotes the mixed morphology state}.}\label{figure4}
\end{center}
\end{figure}


\section{Energy considerations}

The previous analytical results characterize the wetting length in the column state as well as the transition between column and {mixed morphologies}. However, it does not explain the morphology of the {mixed morphology state} and the length of the remaining column. We now compare the surface energy of the different possible shapes and extract the minimal energy to obtain the most stable state. {This method has been successfully used to characterize shape transitions of a drop on chemical stripes \cite{add1,add2,add3} or at topographic steps \cite{add4,add5}}. 

The total surface energy associated with a perfectly wetting liquid lying on the fibers is \cite{protiere2013}
\begin{equation}
\tilde{E}=\frac{E}{\gamma\,a^2}=\tilde{A}_{LV}-\tilde{A}_{SL},
\end{equation}
where $\tilde{E}=E/\gamma\,a^2$ is the dimensionless total energy and $\tilde{A}_{LV}$, $\tilde{A}_{SL}$ are the dimensionless liquid-air and fiber-liquid surface areas normalized by $a^2$, respectively. For the column shape, the surface energy can be expressed as
\begin{equation}\label{Ecolumn}
\tilde{E}_{col}=\int_{-L/a}^{L/a}\left[4\,\frac{R(y)}{a^2}\left(\frac{\pi}{2}-\alpha(y)\right)+\frac{4\,\alpha(y)}{a}\right]\,{\text{d}{y}}.
\end{equation}
The integration takes place over the entire length of the liquid column. The first term is the contribution from the liquid-air interface and the second term is the contribution from the fiber-liquid interface. Note that this formulation allow us  to recover the expression for two parallel fibers ($\alpha$ and ${R}$ are constant all along the column) \cite{protiere2013}.
In the situation studied here, we evaluate (\ref{Ecolumn}) numerically using the expressions for $\alpha$ and $R$ obtained previously for varying distance $d(y)$ between the fibers.

The drop shape is more complex to estimate as no analytical expression of the morphology has been obtained. Using the same argument as \cite{protiere2013}, we estimate the energy of the drop morphology as a sphere of equivalent radius $(3\,\tilde{V}/4\,\pi)^{1/3}$ pierced by two fibers. The  analytical values of the boundary predicted with a sphere model are adjusted as the experimental drop shape is not spherical and better quantitative agreement can be obtained by a value of the energy close to a half-sphere \cite{protiere2013}. These steps lead to the energy for the drop:
{\begin{equation} \label{E_drop}
\tilde{E}_{drop}=0.6\!\!\left[(36\,\pi)^{1/3}\!\left(\frac{V}{a^3}\right)^{2/3}\!\!\!-\pi\,\sqrt{\left(\frac{6\,{V}}{\pi\,a^3}\right)^{2/3}\!\!\!\!-4\,\frac{{d(L)}^2}{a^2}}\right]
\end{equation}}

\noindent We also need to use the constraint between the liquid volume and the length of the liquid column $L$. 

The results of our investigation of the surface energy for the different morphologies are presented in Fig. \ref{figure4}. We first observe that in the {mixed morphology state}, the configuration with one drop at one end is more favorable than two drops, one at both ends, or a single drop centered at the node, which is in agreement with experimental observations. Our results also show that the most stable state in the {mixed morphology state} is obtained for a column that spreads until $d/a \simeq 0.8$ after which one drop is set for $d/a=0.8$ (Fig. \ref{figure4}(a)). In addition, we report the minimum value of this surface energy for $V/a^3=1000$ and increasing tilting angle $\delta$, and we obtain different equilibrium shapes (Fig. \ref{figure4}(b)). More precisely, if the angle is small enough such that all of the fluid can spread in a column, the column morphology is the most stable as observed experimentally. Then for larger angles, the predicted morphology is the {mixed morphology regime} with one drop until the angle $\delta$ reaches a critical value where the preferred state becomes one drop centered at the nodes. This diagram captures quantitatively the transition between the column and {mixed morphology} state. However, the analytical determination of the symmetry breaking transition to a centered drop is only qualitative as the transition is highly sensitive to the prefactor used in (\ref{E_drop}).

\begin{figure}
\begin{center}
\includegraphics[width=8.8cm]{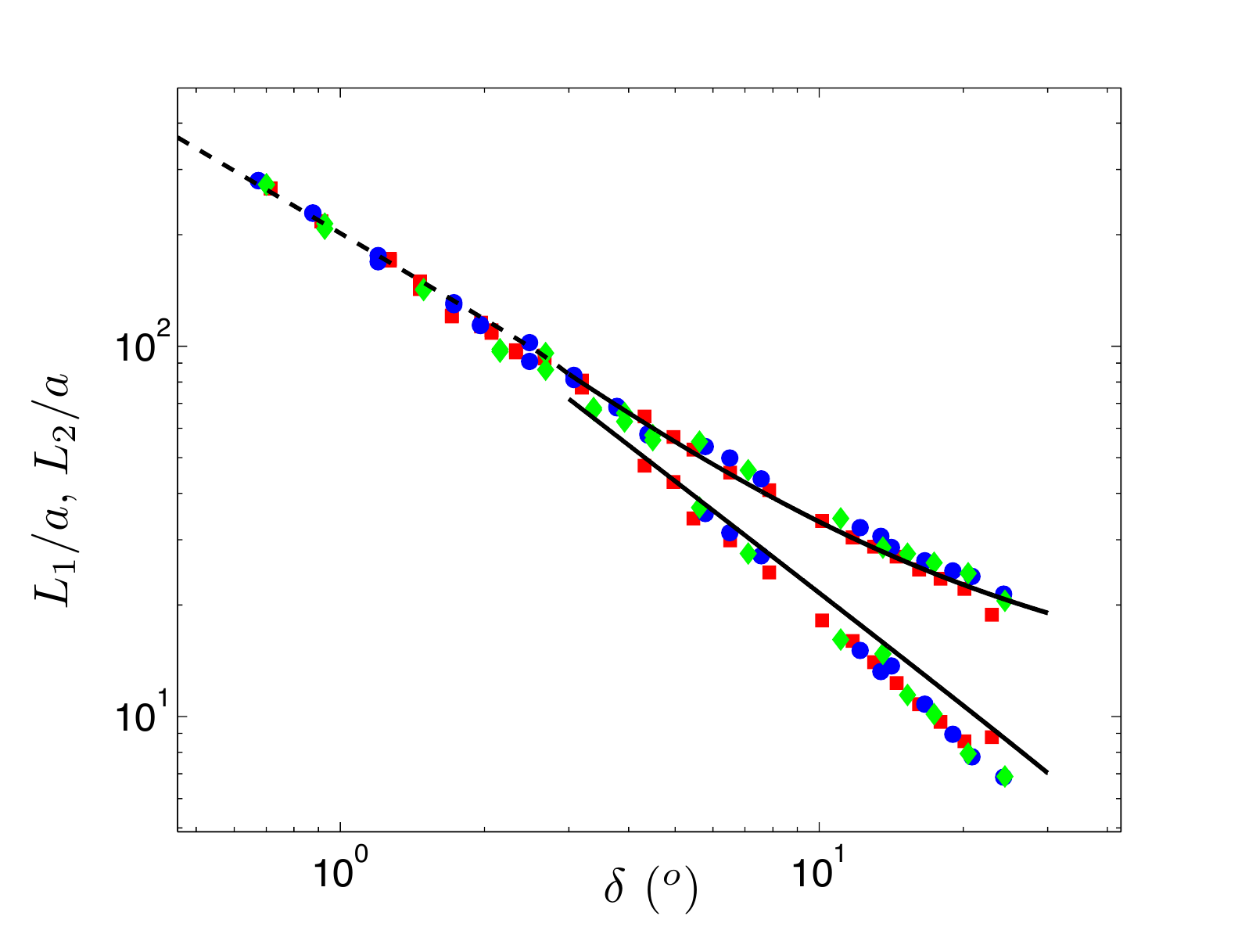}
\caption{Evolution of the wetting length of the drop on both sides of the nodes, $L_1/a$ and $L_2/a$, with varying tilting angle $\delta$ for nylon fibers of diameter $2\,a=0.245$ mm (\textcolor{blue}{$\bullet$}), $2\,a=0.30$ mm (\textcolor{green}{$\blacklozenge$})  and $2\,a=0.35$ mm (\textcolor{red}{$\blacksquare$}) for $V/a^3=1000$. The dashed-dotted line is the analytical prediction in the column state and the continuous line is the prediction in the {mixed morphology state.}}\label{figure5}
\end{center}
\end{figure}

Finally, using the value of $d/a$ that minimizes the surface energy (inset of Fig. \ref{figure4}b), we can compare the predicted length on both side of the nodes for increasing angle. These results are reported in Fig. \ref{figure5} and capture well the experimental measurements. At sufficiently small $\delta$, the liquid spreads in a column state symmetrically with respect to the nodes. Then, at larger $\delta$, the transition to a non-symmetric state can be predicted by considering on one side a liquid column of extension satisfying $d/a=d_{\rm min}/a$ (see Fig. \ref{figure4}(b)). On the other side, a liquid column of the same extension is considered but we added the presence of a drop that contains the remaining liquid.


\section{Conclusion}

In conclusion, in this Letter we have described the morphology of a perfectly wetting fluid lying at the node of two crossed fibers. The morphology depends on the tilt angle as well as the volume of liquid and range from a liquid column to a drop but can also exhibit a non-symmetric {mixed morphology state} that is not observed with parallel fibers \cite{protiere2013}. An analytical model allows us to capture both the transition between these different regimes and the shape of the liquid such as the extension of the column. One prediction of the present study is that shearing a material consisting of random orientation of fibers will decrease the average angle between the fibers and so tune the liquid morphology. Thus, the transition from a drop to a column state leads to a larger area at the liquid/air interface, which would enhance the evaporation rate. Consequently, this idea suggests that in a fibrous media the drying efficiency can be controlled mechanically.

\acknowledgments
We thank Unilever Research and NSF CBET-1132835 for support of this research. We thank P. Warren and S. Protiere for helpful discussions.

\bibliographystyle{eplbib}

\begin{thebibliography}{10}
\expandafter\ifx\csname url\endcsname\relax\def\url#1{\texttt{#1}}\fi

\bibitem{last}
\Name{Liew T.~P. \and Conder J.~R.} \REVIEW{J. Aerosol Sci.}{16}{1985}{497}.

\bibitem{contal2004}
\Name{Contal P., Simao J., Thomas D., Frising T., Call{\'e} S., Appert-Collin
  J. \and B{\'e}mer D.} \REVIEW{J. Aerosol Sci.}{35}{2004}{263}.

\bibitem{hubbe2008}
\Name{Hubbe M.~A., Rojas O.~J., Lucia L.~A. \and Sain M.}
  \REVIEW{BioResources}{3}{2008}{929}.

\bibitem{stone2001}
\Name{Schleier-Smith J.~M. \and Stone H.~A.} \REVIEW{Phys. Rev.
  Lett.}{86}{2001}{3016}.

\bibitem{tregzes2002}
\Name{Tegzes P., Vicsek T. \and Schiffer P.} \REVIEW{Phys. Rev.
  Lett.}{89}{2002}{094301}.

\bibitem{nowak2005}
\Name{Nowak S., Samadani A. \and Kudrolli A.} \REVIEW{Nat. Phys.}{1}{2005}{50}.

\bibitem{herminghaus2005}
\Name{Herminghaus S.} \REVIEW{Adv. Phys.}{54}{2005}{221}.

\bibitem{kudrolli2008}
\Name{Kudrolli A.} \REVIEW{Nat. Mat.}{7}{2008}{174}.

\bibitem{chopin2011}
\Name{Chopin J. \and Kudrolli A.} \REVIEW{Phys. Rev. Lett.}{107}{2011}{208304}.

\bibitem{bonn2012}
\Name{Pakpour M., Habibi M., Moller P. \and Bonn D.} \REVIEW{Sci.
  Rep.}{2}{2012}{549}.

\bibitem{strauch2012}
\Name{Strauch S. \and Herminghaus S.} \REVIEW{Soft Matter}{8}{2012}{8271}.

\bibitem{willett2000}
\Name{Willett C.~D., Adams M.~J., Johnson S.~A. \and Seville J.~P.}
  \REVIEW{Langmuir}{16}{2000}{9396}.

\bibitem{rod1}
\Name{B\"{o}rzs\"{o}nyi T., Szab\'o B., T\"{o}r\"{o}s G., Wegner S.,
  T\"{o}r\"{o}k J., Somfai E., Bien T. \and Stannarius R.} \REVIEW{Phys. Rev.
  Lett.}{108}{2012}{228302}.

\bibitem{rod2}
\Name{Gravish N., Franklin S.~V., Hu D.~L. \and Goldman D.~I.} \REVIEW{Phys.
  Rev. Lett.}{108}{2012}{208001}.

\bibitem{rod3}
\Name{Guo Y., Wassgren C., Hancock B., Ketterhagen W. \and Curtis J.}
  \REVIEW{Phys. Fluids}{25}{2013}{063304}.

\bibitem{rod4}
\Name{Borzsonyi T. \and Stannarius R.} \REVIEW{Soft Matter}{9}{2013}{7401}.

\bibitem{caroll1976}
\Name{Carroll B.~J.} \REVIEW{J. Colloid Interface Sci.}{57}{1976}{488}.

\bibitem{caroll1986}
\Name{Carroll B.~J.} \REVIEW{Langmuir}{2}{1986}{248}.

\bibitem{chou2011}
\Name{Chou T.~H., Hong S.~J., Liang Y.~E., Tsao H.~K. \and Sheng Y.~J.}
  \REVIEW{Langmuir}{27}{2011}{3685}.

\bibitem{princen1970}
\Name{Princen H.~M.} \REVIEW{J. Colloid Interface Sci.}{34}{1970}{171}.

\bibitem{bedarkar2010}
\Name{Bedarkar A., Wu X.-F. \and Vaynberg A.} \REVIEW{Appl. Surf.
  Sci.}{256}{2010}{7260}.

\bibitem{protiere2013}
\Name{Protiere S., Duprat C. \and Stone H.~A.} \REVIEW{Soft
  Matter}{9}{2013}{271}.

\bibitem{bico2004}
\Name{Bico J., Roman B., Moulin L. \and Boudaoud A.}
  \REVIEW{Nature}{432}{2004}{690}.

\bibitem{duprat2012nature}
\Name{Duprat C., Protiere S., Beebe A.~Y. \and Stone H.~A.}
  \REVIEW{Nature}{482}{2012}{510}.

\bibitem{duprat2013}
\Name{Duprat C., Bick A.~D., Warren P.~B. \and Stone H.~A.}
  \REVIEW{Langmuir}{29}{2013}{7857}.

\bibitem{mullins}
\Name{Mullins B.~J., Agranovski I.~E., Braddock R.~D. \and Ho C.~M.} \REVIEW{J.
  Colloid Interface Sci.}{269}{2004}{449}.

\bibitem{claussen}
\Name{Claussen J.~O.} \Book{Elasticity and {M}orphology of {W}et {F}ibers}
  Ph.D. thesis University of G\"{o}ttingen (2011).

\bibitem{gilet2009}
\Name{Gilet T., Terwagne D. \and Vandewalle N.} \REVIEW{App. Phys.
  Lett.}{95}{2009}{014106}.

\bibitem{quere_conical}
\Name{Lorenceau E. \and Qu\'er\'e D.} \REVIEW{J. Fluid Mech.}{510}{2004}{29}.

\bibitem{add1}
\Name{Brinkmann M. \and Lipowsky R.} \REVIEW{J. Appl. Phys.}{92}{2002}{4296}.

\bibitem{add2}
\Name{Gau H., Herminghaus S., Lenz P. \and Lipowsky R.}
  \REVIEW{Science}{283}{1999}{46}.

\bibitem{add3}
\Name{Ferraro D., Semprebon C., T—th T., Locatelli E., Pierno M., Mistura G.
  \and Brinkmann M.} \REVIEW{Langmuir}{28}{2012}{13919}.

\bibitem{add4}
\Name{Brinkmann M. \and Blossey R.} \REVIEW{Eur. Phys. J. E}{14}{2004}{79}.

\bibitem{add5}
\Name{Kusumaatmaja H., Lipowsky R., Jin C., Mutihac R.~C. \and Riegler H.}
  \REVIEW{Phys. Rev. Lett.}{108}{2012}{126102}.

\end{thebibliography}

\end{document}